\documentclass[12pt]{article}
\usepackage{amssymb,graphicx,url}
\usepackage{epstopdf}
\usepackage{amsmath,amsfonts}
\usepackage{epsfig,color}

\begin{document}

\vskip -0.9cm

\sloppy

\title{\bf Initial Conditions for Imperfect Dark Matter}
\author{Sabir Ramazanov\footnote{{\bf e-mail}:
    Sabir.Ramazanov@ulb.ac.be}\\
 \small{\em Service de Physique Th\'eorique, Universit\' e  Libre de Bruxelles (ULB),}\\
\small{\em CP225 Boulevard du Triomphe, B-1050 Bruxelles, Belgium}
}
{\let\newpage\relax\maketitle}


\begin{abstract}
We discuss initial conditions for the recently proposed Imperfect Dark Matter (Modified Dust). We 
show that they are adiabatic under fairly moderate assumptions about the cosmological evolution of the 
Universe at the relevant times. 
\end{abstract}

\section{Introduction and Summary}

Planck data favors adiabatic initial conditions at the onset of the 
hot Big Bang~\cite{Ade:2015xua}. This corresponds to the situation in the early Universe, 
when the number densities of different particle species, e.g., dark matter (DM) particles 
and photons, are universally distributed in space. That picture, comfortably 
accomodated in the $\Lambda \mbox{CDM}$ cosmology supplemented by the short stage 
of an inflationary expansion, can be less transparent 
in more exotic setups. In particular, adiabaticity of initial conditions 
is obscured in the case, if DM has a non-particle origin.

In the present paper, we continue to discuss Imperfect Dark Matter (IDM) scenario~\cite{Capela:2014xta, Mirzagholi:2014ifa} 
(Modified Dust in Ref.~\cite{Capela:2014xta}). The action of IDM is given by   
\begin{equation}
\label{action}
S=\int d^4 x \sqrt{-g} \left[\frac{\lambda}{2} \left( g^{\mu \nu} 
\partial_{\mu} \varphi \partial_{\nu} \varphi-1  \right) +\frac{\gamma (\varphi)}{2} (\square \varphi)^2\right]\; ,
\end{equation}
(see Ref.~\cite{Lim:2010yk} for generalizations). Here $\lambda$ and $\varphi$ are two scalars, and $\gamma (\varphi)$ is some function of the 
field $\varphi$. The Lagrange multiplier $\lambda$ enforces the constraint $g^{\mu \nu} \partial_{\mu} \varphi \partial_{\nu} \varphi=1$. This equation
defines the geodesics followed by the collisionless particles, with the field $\varphi$ being the velocity 
potential. The analogy with the pressureless perfect fluid, dust, is complete in the situation when the higher derivative (HD) term is absent, 
i.e., $\gamma (\varphi) =0$. In that case, the Lagrange multiplier $\lambda$ plays the 
role of the energy density of dust: it redshifts away as the inverse 
third power of the scale factor $a$ in the expanding Universe. While the dust provides a good description of DM 
to the linear level, it should be modified in the non-linear regime. The reason is that 
it develops caustic singularities, i.e., the physical quantities,---the velocity dispersion and 
the energy density,---blow up at the finite time~\cite{Landau, Sahni:1995rm}. This drawback of the dust model was one of the motivations (not the unique one, though) to introduce the 
HD term~\cite{Capela:2014xta}. Before we summarize some of the effects arising due to the non-zero 
function $\gamma (\varphi)$, let us discuss, how the action~\eqref{action} emerges in different physical frameworks. 

The mimetic dark matter scenario~\cite{Chamseddine:2013kea, Chamseddine:2014vna} deals with the conformally transformed metric 
of the form\footnote{The generalization of mimetic dark matter scenario to disformal metrics transformations has been constructed in Ref.~\cite{Deruelle:2014eha}.},
\begin{equation}
\nonumber
g_{\mu \nu} =\left(\tilde{g}^{\alpha \beta}\partial_{\alpha} \varphi \partial_{\beta} \varphi\right) 
\tilde{g}_{\mu \nu} \; .
\end{equation} 
Rather unexpectedly, variation of the general relativity action with respect to the auxilliary metric 
$\tilde{g}_{\alpha \beta}$ and the scalar $\varphi$, gives more 
than just Einstein's equations. The difference is about a new degree 
of freedom behaving as dust. This returns us to the action~\eqref{action} up to the 
HD term~\cite{Golovnev:2013jxa}, which is set by hands in this context~\cite{Chamseddine:2014vna}. Alternatively, one can trace back the origin of IDM to Lorentz 
violating theories of gravity. In particular, 
the action~\eqref{action} arises in the infrared 
limit of the Horava--Lifshitz model with the projectability 
condition applied~\cite{Horava:2009uw, Mukohyama:2009mz, Blas:2009yd}. More generically, IDM is closely related to a version of 
Einstein--Aether 
theory~\cite{Jacobson:2000xp} with the aether 
field being the derivative of the scalar~\cite{Haghani:2014ita, Jacobson:2014mda} (see also Refs.~\cite{Haghani:2015iva, Jacobson:2015mra, Speranza} for the most recent discussion on the topic). 

During the large part of IDM evolution, one assumes that the function $\gamma (\varphi)$ is constant. In this situation, 
the model possesses shift symmetry $\varphi \rightarrow \varphi + c$. There is, consequently, the Noether charge density that redshifts 
away as the inverse third power of the scale factor. Up to the term suppressed by the $\gamma$ factor, 
the energy density of IDM is equal to the Noether 
charge density~\cite{Mirzagholi:2014ifa} (Section~2). Hence, with a good accuracy, the cosmological evolution of IDM mimics that of 
the dust particles \footnote{This is an exact statement in 
the situation, when IDM is the only component of the Universe. See the discussion in Section~2}. The degeneracy gets broken at the linear level, where
 the $\gamma$-term leads to the constant sound speed $c^2_s \simeq \gamma$~\cite{Chamseddine:2014vna}. 
This sets a cutoff on the power spectrum at sufficiently small scales: beyond the 
sound horizon energy density perturbations do not grow. Therefore, they are suppressed 
compared to the predictions of cold dark matter (CDM) scenarios. In particular, setting \footnote{Note that we use the 
convention $8 \pi M^2_{Pl}=1$ throughout the paper, where $M_{Pl}$ is the Planck mass. An associated value of the parameter $\gamma$ 
reads in units of Ref.~\cite{Capela:2014xta} $\gamma \sim 10^{-10} M^2_{Pl}$.} $\gamma \sim 10^{-9}$, one can suppress the growth of structures with the comoving wavelength $\lesssim 100~\mbox{kpc}$, alleviating the mismatch between the observed number 
of dwarf galaxies and the value predicted in the CDM framework~\cite{Klypin:1999uc,Moore, Weinberg:2013aya}. Alternatively, however, the small 
scale problems may be the consequence of the incorrect implementation of several baryonic processes, as 
indicated most recently in the analysis of Ref.~\cite{Sawala:2014xka}. In that case, 
one rather deals with the constraint~\cite{Capela:2014xta} 
\begin{equation}
\label{constraint}
\gamma \lesssim 10^{-9} \; . 
\end{equation}
For those small values of the parameter $\gamma$, the linear evolution of IDM perturbations is analogous to that of 
CDM given that they start from the same initial conditions (see the discussion below). The important difference, 
however, may arise at the structure formation level, i.e., in the non-linear regime~\cite{Capela:2014xta}. This line of discussion is 
far out of the scope of the present paper. Here we will be interested in the opposite situation, namely, 
when the relevant cosmological modes are in the deep super-horizon regime. 

If the shift symmetry is exact at {\it all} the times, 
IDM cannot be the main component of the invisible matter in the 
Universe. The reason is that the Noether charge density gets washed 
out during inflation with an exponential accuracy.   
In this situation, IDM constitutes only a tiny, ${\cal O}(\gamma)$, fraction of the overall DM during the dust dominated stage~\cite{Mirzagholi:2014ifa}. 
To avoid this, 
one necessarily assumes breaking of the shift symmetry taking place at the early stages of the 
Universe, i.e., deeply in the radiation dominated (RD) era. In IDM model, this is realized by promoting the constant $\gamma$ 
to the function $\gamma (\varphi)$\footnote{Another idea would be to couple the field $\varphi$ to the inflaton~\cite{Chamseddine:2013kea}. 
In the present paper, however, we consider only gravitational interactions between IDM and other fields.}. We assume furthermore that the variation of the function $\gamma (\varphi)$ 
is substantial only at very early times, and negligible otherwise. In that way, one can easily generate the amount of 
the Noether charge required to explain cosmological experiments, as we review in Section~2. Notably, perturbations of IDM produced 
by the same mechanism are adiabatic with a high accuracy. 

Before the Noether charge gets produced, IDM tracks the dominant matter of the Universe, 
e.g., it has an equation of state of radiation in the Universe driven by the relativistic particles. 
At the linear level, the similarity with the dominant matter reveals in an exact 
adiabaticity of IDM perturbations (Section~3)\footnote{The analogous observation has been made in the earlier work~\cite{Capela:2014xta}. Here we show 
that this is an exact statement. We also generalize it to any dominant matter in the Universe, while the case of radiation has been considered in 
Ref.~\cite{Capela:2014xta}.}. That situation is quite analogous to what 
one has in the $\Lambda \mbox{CDM}$ cosmology at very high temperatures, when all the 
particle species are in the thermodynamic equilibrium. 
More importantly, the adiabaticity of IDM perturbations holds later on, after the shift-symmetry breaking occurs, despite the presence of the non-adiabatic pressure. We prove this by the explicit computation of the curvature 
perturbation of IDM, i.e., $\zeta_{IDM}$. We show that the latter corresponds to the adiabatic initial 
conditions for IDM, at least under rather moderate assumptions: the variation of the gravitational potential must be negligible 
relatively to the Hubble rate at the times, when the shift-symmetry breaking takes place. That condition is obeyed 
with a high accuracy well before the matter/radiation equality. In particular, this guarantees that IDM is indistinguishable from  CDM at the level of cosmic 
microwave background measurements\footnote{This way of setting initial conditions is to be compared to that of Ref.~\cite{Capela:2014xta}. 
In the latter paper, IDM perturbations relied on the arbitrary constant of integration. That is, the adiabaticity of initial conditions in Ref.~\cite{Capela:2014xta} 
was at the price of tuning an arbitrary constant to some particular value. Alternatively, any other choice would result in an admixture 
of an isocurvature mode.}.

The outline of the paper is as follows. In Section~2, we briefly review the IDM 
scenario including the mechanism for producing the Noether charge density in that picture. 
In Section~3, we discuss the super-horizon evolution of IDM perturbations and show that they 
are adiabatic under rather general conditions.

\section{Generating dark matter}
We start with writing down the system of equations following from the 
action~\eqref{action}. The simplest one is the constraint 
\begin{equation}
\label{geodesics}
g^{\mu \nu} \partial_{\mu} \varphi \partial_{\nu} \varphi =1 \; ,
\end{equation}
obtained from the variation with respect to the Lagrange multiplier. 
Applying the covariant derivative to the constraint~\eqref{geodesics}, 
one obtains the geodesic equation. In this regard, IDM is equivalent to the 
collection of dust particles moving in the gravitational field. This degeneracy, however, 
gets broken at the level of the other equations. 
Variation of the action~\eqref{action} with respect to the field $\varphi$ yields~\cite{Mirzagholi:2014ifa} 
\begin{equation}
\label{noncons}
\nabla_{\mu} J^{\mu} =\frac{1}{2} \gamma_{\varphi} (\square \varphi)^2 \; ,
\end{equation}
where 
\begin{equation}
\label{current}
J^{\mu} =(\lambda -\gamma_{\varphi} \square \varphi )\partial^{\mu} \varphi -\gamma (\varphi) \partial^{\mu} \square \varphi \; .
\end{equation}
Finally, the energy-momentum tensor corresponding to the action~\eqref{action} is given by~\cite{Mirzagholi:2014ifa, Chamseddine:2014vna}
\begin{equation}
\nonumber
\begin{split}
T^{\mu}_{\nu} &=\Bigl(\lambda -2\gamma_{\varphi} (\varphi) \square \varphi \Bigr)\partial_{\nu} \varphi \partial^{\mu} \varphi 
+\gamma (\varphi)\Bigl(\partial_{\alpha} \varphi \partial^{\alpha} \square \varphi +
\frac{1}{2} (\square \varphi)^2 +\frac{\gamma_{\varphi} (\varphi)}{\gamma (\varphi)} \square \varphi \Bigr) \delta^{\mu}_{\nu}-\\
& -\gamma (\varphi) \left(\partial_{\nu} \varphi \partial^{\mu} \square \varphi 
+\partial_{\nu} \square \varphi \partial^{\mu} \varphi \right) \; .
\end{split}
\end{equation}
In the present paper we are mostly interested in the background evolution 
of IDM as well as in the linear evolution of its super-horizon perturbations. That is, 
we will always neglect spatial derivatives of the fields. In this approximation, 
Eq.~\eqref{noncons} takes the form\footnote{We stick to the notion "Noether charge", that 
can be misleading at the times, when the shift symmetry is broken. Hopefully, this is not going 
to confuse the reader.},
\begin{equation}
\label{nonconsnoether}
\frac{d}{dt} \left (\sqrt{-g} n \right) =\frac{1}{2} \sqrt{-g} \gamma_{\varphi} (\varphi) (\square \varphi)^2  \; ,
\end{equation}
where $n \equiv J^0$ is the Noether charge density. We assume that during most of the stages of the evolution, the function $\gamma (\varphi)$ is constant, i.e., the model is shift symmetric. 
In that case, the r.h.s. of Eq.~\eqref{noncons} is zero and one has
\begin{equation}
\label{noethercons}
\frac{d (na^3)}{dt} =0 \qquad n=\lambda -3\gamma \dot{H} \; .  
\end{equation}
As it follows, the Noether charge density redshifts away as $1/a^3$, i.e., $n=C/a^3$. 
Assuming that IDM obeys shift symmetry during the inflationary expansion of the Universe, the
 density $n$ relaxes to zero with an exponential accuracy, i.e., $n=0$ 
at the onset of the hot Big Bang. Otherwise, one would need to set an extremely large value of $n$ in the beginning of inflation. 
 
 In the situation with the zeroth 
Noether charge, IDM tracks the total matter filling in the Universe~\cite{Mirzagholi:2014ifa}. 
Indeed, the physical energy density is related to the Noether charge density by  
\begin{equation}
\label{epsnoether}
\rho_{IDM} \equiv T^{0}_{0}=n +\frac{3\gamma}{2} \rho_{tot}  \; ,
\end{equation}
while the pressure is given by 
\begin{equation}
\label{pressure}
{\cal P}_{IDM}\equiv -\frac{1}{3}T^{i}_{i}=\frac{3\gamma}{2} {\cal P}_{tot}-3\dot{\gamma} H \; .
\end{equation}
Here we made use of equations $H^2 =\frac{1}{3} \rho_{tot}$ and $-2\dot{H}-3H^2={\cal P}_{tot}$; 
the subscript '$tot$' stands for the total matter including IDM. 
Neglecting the Noether charge density as well as the variation of the parameter $\gamma$ in Eqs.~\eqref{epsnoether} and~\eqref{pressure}, 
one obtains $w_{IDM} \equiv \frac{{\cal P}_{IDM}}{\rho_{IDM}}=w_{tot}$, where $w_{tot} \equiv \frac{{\cal P}_{tot}}{\rho_{tot}}$. This proves the point made above.

For the future convenience, we prefer to rephrase Eqs.~\eqref{epsnoether} and~\eqref{pressure} by splitting the contributions to the total energy density and pressure, which follow from IDM and the external 
matter fields. This gives
\begin{equation} 
\label{epsnoethernew}
\rho_{IDM} =\frac{2n}{2-3 \gamma} +\frac{3\gamma}{2-3\gamma} \rho_{ext} 
\end{equation}
and 
\begin{equation}
\label{pressurenew}
{\cal P}_{IDM}=\frac{3\gamma}{2-3\gamma} {\cal P}_{ext}-6\frac{\dot{\gamma}}{2-3\gamma} H \; .
\end{equation}
Here $\rho_{ext} \equiv  \rho_{tot} -\rho_{IDM}$ and ${\cal P}_{ext} \equiv {\cal P}_{tot}-{\cal P}_{IDM}$ 
denote the total energy density and pressure of all the matter fields not including 
IDM, respectively. As it follows from Eqs.~\eqref{epsnoethernew} and~\eqref{pressurenew}, IDM behaves as a perfect tracker also with respect to the external matter. Namely, setting the Noether charge density and the 
derivative of the parameter $\gamma$ to zero, we have $w_{IDM}=w_{ext}$, where $w_{ext} \equiv \frac{{\cal P}_{ext}}{{\cal \rho}_{ext}}$. In particular, this means that for a very small value of the parameter $\gamma$, IDM may constitute only a tiny fraction of the 
invisible matter during the dust dominated stage. We are interested in the different opportunity of IDM 
being the main (the only) component of DM.

Since this point on and until the end of the Section, we assume that there is a period of shift-symmetry breaking taking place deeply in the RD stage. 
This allows to generate non-zero Noether charge, which, by the suitable choice of parameters, 
can be tuned to match the observed value~\cite{Mirzagholi:2014ifa}. 
With no much loss of generality, we consider an instantaneous 
transition from some initial value $\gamma_1$ to the value $\gamma_2 >\gamma_1$. Namely, the function $\gamma (\varphi)$ 
has the form
\begin{equation}
\label{inst}
\gamma (\varphi)=\gamma_1 +\Delta \gamma\sigma (\varphi-\varphi_*) \; ,
\end{equation}
where $\sigma (\varphi-\varphi_*)$ is the Heaviside function; $\varphi_*$ is the constant 
pointing the time, when the transition happens, and $\Delta \gamma \equiv \gamma_2-\gamma_1$.  

Accordingly, the value of the Noether charge undergoes an instantaneous flip. Integrating Eq.~\eqref{nonconsnoether} with the initial 
condition $n=0$ set at $t<t_*$, we obtain 
\begin{equation}
\label{back}
na^3 =  \frac{9}{2} \Delta \gamma \cdot H^2 a^3 \left. \right|_{t=t_*} \quad t>t_* . 
\end{equation} 
One can tune the time and the quantity $\Delta \gamma$, so that 
we get the correct energy density of DM today. As it follows, earlier the Noether charge is generated, less value of $\Delta \gamma$ is required. In particular, 
for the choice $\Delta \gamma \lesssim \gamma \lesssim 10^{-9}$, this indeed happens at very early times. The corresponding redshifts and temperatures read $z \gtrsim 10^{12}-10^{13}$ and $T \gtrsim 100~\mbox{MeV}-1~\mbox{GeV}$, respectively. Soon after the Noether charge is 
produced, it becomes the total contribution to the energy density of IDM. Indeed, 
the second term in Eq.~\eqref{epsnoethernew} mimicking the behaviour of the external matter of the Universe 
(radiation) redshifts away fast relatively to the Noether charge density. Since this point on, the cosmological evolution of IDM resembles that of the 
dust particles. This is an exact statement in the situation, when IDM is the only matter 
in the Universe. In that case, $\rho_{IDM}=\rho_{tot}$ and ${\cal P}_{IDM}={\cal P}_{tot}$. 
From Eqs.~\eqref{epsnoether} and~\eqref{pressure}, one then easily obtains ${\cal \rho}_{IDM} \propto n \propto 1/a^3$ and 
${\cal P}_{IDM}=0$~\cite{Capela:2014xta, Mirzagholi:2014ifa, Haghani:2014ita}.

In the remainder of the paper, we show that perturbations of IDM generated 
by the same mechanism are adiabatic under fairly relaxed assumptions.

\section{Super-horizon evolution of IDM perturbations}
\subsection{Generalities}

Before we dig into the details of the linear level analysis, let us specify the gauge choice. 
Generically, the metric reads to the linear order, 
\begin{equation}
\nonumber 
ds^2 =(1+2\Phi) (dx^0)^2 +2a\partial_i Z dx^0 dx^i-a^2 (\delta_{ij}+2\Psi \delta_{ij}-\partial_i \partial_j E) dx^i dx^j \; ,
\end{equation}
where $E$, $Z$, $\Phi$ and $\Psi$ are the scalar potentials, and we omitted vector and tensor perturbations. See Refs.~\cite{Mukhanov:1990me, Mukhanov, Gorbunov:2011zzc} for the reviews and textbooks on the 
cosmological perturbation theory. Calculations 
look particularly simple and transparent in the synchronous 
gauge, which we use in the bulk of the paper. In the Appendix~A, we provide calculations in the 
Newton's gauge in order to cross-check the results.

To go from the Newton's gauge, where $Z=0$ and $E=0$, to the synchronous gauge, one makes the coordinate transformation, 
$\tilde{x}_{\mu}=x_{\mu}+\xi_{\mu}$, where $\xi_0 =\delta \varphi$ and $\xi_i$ obey the condition 
$\partial_0 \xi_i +\partial_i \xi_0=0$. With this coordinate choice, the perturbation $\delta \varphi$ 
and the potential $\Phi$ turn into zero,
\begin{equation}
\nonumber
\delta \varphi \rightarrow \delta \tilde{\varphi}=0 \qquad \Phi \rightarrow \tilde{\Phi}=0 \; .
\end{equation}
The fact that perturbations of the field $\varphi$ and the potential $\Phi$ can be switched to zero 
simultaneously follows from the gauge-independent relation $\delta \dot{\varphi} =\Phi$,---the constraint~\eqref{geodesics} linearized. In the synchronous gauge, the potential $\Psi$ is given by
\begin{equation}
\label{relation}
\Psi \rightarrow \tilde{\Psi}=\Psi -H \delta \varphi \; .
\end{equation}
The condition $\partial_i \xi_0 +\partial_0 \xi_i=0$ then guarantees that the $(0i)$-component of the metric remains zero. At the same time, 
the potential $E$ is generically non-zero in the synchronous gauge. This is, however, negligible in the super-horizon regime, 
which is the case of our primary interest. Indeed, $\partial_i \partial_j\tilde{E} = 
-(\partial_i \xi_j +\partial_j \xi_i)/a^2 \propto \partial_i \partial_j \delta \varphi /(a^2H) \rightarrow 0$. 

Hereafter, we prefer to omit the tilde over the transformed quantities. In the synchronous gauge, the Noether charge density perturbation is 
given by (see Eq.~\eqref{current}),
\begin{equation}
\label{linear}
 \delta n =
\delta \lambda -3\gamma \ddot{\Psi}  -3\dot{\gamma} \dot{\Psi}\; .
\end{equation}
In what follows, we will also need the expressions for IDM energy density and pressure perturbations. 
These are given by  
\begin{equation}
\label{endens}
\delta \rho_{IDM}=\delta n +9\gamma H \dot{\Psi} 
\end{equation}
and
\begin{equation}
\label{presspert} 
\delta {\cal P}_{IDM}=-9 \gamma H \dot{\Psi} -3\gamma \ddot{\Psi} -3\dot{\gamma} \dot{\Psi} \; ,
\end{equation}
respectively. 

\subsection{Before shift-symmetry breaking}
We first discuss the case of the zeroth Noether charge. Let us show that perturbations of IDM are exactly adiabatic in that case. One obtains from Eq.~\eqref{endens},
\begin{equation}
\label{energybefore}
\delta \rho_{IDM}=9\gamma H \dot{\Psi} \; .
\end{equation}
The analogous expression for the energy density perturbation of the external matter can 
be inferred from the $00$-th component of Einstein's equations, which reads in the 
synchronous gauge
\begin{equation}
\label{00}
3H\dot{\Psi}=\frac{1}{2}{\delta \rho}_{ext}+\frac{1}{2} {\delta \rho}_{IDM} \; .
\end{equation}
Combining this with Eq.~\eqref{energybefore}, one obtains 
\begin{equation}
\nonumber
\delta \rho_{IDM}= \frac{3\gamma}{2-3\gamma} \delta \rho_{ext} \; .
\end{equation}
We plug the latter into the definition of the curvature perturbation of IDM, 
\begin{equation}
\label{zeta} 
\zeta_{IDM} =\Psi +\frac{\delta \rho_{IDM}}{3(\rho_{IDM}+{\cal P}_{IDM})} \; ,
\end{equation}
and then make use of the relation
\begin{equation}
\nonumber
\rho_{IDM}+{\cal P}_{IDM}=\frac{3\gamma}{2-3\gamma} (\rho_{ext}+{\cal P}_{ext}) \; .
\end{equation}
This expression follows from Eqs.~\eqref{epsnoethernew} and~\eqref{pressurenew}, where one should set the Noether charge density to zero and 
the parameter $\gamma$ to a constant. We obtain,
\begin{equation}
\label{adia}
\zeta_{IDM} =\Psi + \frac{\delta \rho_{ext}}{3(\rho_{ext}+{\cal P}_{ext})} \equiv \zeta_{ext} \; ,
\end{equation}
where $\zeta_{ext}$ is the curvature perturbation of the external matter. 
Finally, we use the relation between the partial curvature perturbations 
corresponding to the external matter fields and the quantity $\zeta_{ext}$, 
\begin{equation}
\nonumber 
\zeta_{ext} =\sum_{i \neq IDM} \frac{\dot{\rho}_{i}}{\dot{\rho}_{ext}}\zeta_{i} \; .
\end{equation}
Here the index $i $ stands for the particular matter field, i.e., photons, neutrinos, baryons etc., 
and the subscript $i \neq \mbox{IDM}$ means that the contribution of IDM has been omitted. 
Under an assumption that there is no admixture of the baryon or neutrino isocurvature modes, 
all the partial curvature perturbations $\zeta_{i \neq IDM}$ are equal between each other. 
Since an equality $\sum_{i \neq IDM} \dot{\rho}_i /\dot{\rho}_{ext}=1$, one has $\zeta_{i \neq IDM} =\zeta_{ext}$. Comparing this with Eq.~\eqref{adia}, we get
\begin{equation}
\label{adiabat} 
\zeta_{IDM}=\zeta_{i \neq IDM} \; .
\end{equation}
That is, the IDM curvature perturbation equals to the curvature perturbations of 
the standard matter fields. This means that IDM perturbations are exactly adiabatic in the shift-symmetric case/before the 
shift-symmetry breaking takes place. That situation is quite similar to what 
happens in the standard cosmology with the particle DM: at very early times all the particle 
species behave as the single fluid, and initial scalar perturbations are adiabatic 
by definition. 

\subsection{After shift-symmetry breaking}
While IDM perturbations are exactly adiabatic before shift-symmetry breaking, it is not immediately clear that they remain so at later 
times. Indeed, starting from initial conditions~\eqref{adiabat}, they can change due to the presence of the non-adiabatic pressure, 
\begin{equation}
\label{nonadpr} 
{\cal P}^{non-ad}_{IDM} \equiv \delta {\cal P}_{IDM}- \frac{\dot{{\cal P}}_{IDM}}{\dot{\rho}_{IDM}}\delta \rho_{IDM} \; ,
\end{equation}
which is not manifestly zero. This is one distinction of the IDM fluid from the familiar fluids, i.e., radiation and dust. 
Consequently, the IDM curvature perturbation evolves 
behind the horizon accordingly to the equation~\cite{Mukhanov:1990me}, 
\begin{equation}
\label{evolution}
\dot{\zeta}_{IDM} =-\frac{H}{\rho_{IDM} +{\cal P}_{IDM}} {\cal P}^{non-ad}_{IDM} \; ,
\end{equation}
while those of the standard matter fields remain constant. This results into the 
violation of the condition~\eqref{adiabat} leading to the appearance of an isocurvature mode and, consequently, to a potential conflict with 
the Planck data. The non-adiabatic pressure, however, is negligible in two regimes: at the times $t < t_*$ and $t \gg t_*$. 
The former follows from an exact adiabaticity of IDM perturbations at the early times implied by Eq.~\eqref{adiabat}\footnote{Not referring to Eq.~\eqref{adiabat}, 
one can show that the non-adiabatic pressure of IDM at the times $t<t_*$ is proportional to that of the external 
matter, i.e., ${\cal P}^{non-ad}_{IDM} \propto{\cal P}^{non-ad}_{ext}$, where the subscript 'ext' stands for 
the combination of photons, neutrinos, baryons etc. As perturbations of the external matter fields are 
assumed to be adiabatic, one has ${\cal P}^{non-ad}_{ext}=0$. Consequently, ${\cal P}^{non-ad}_{IDM}=0$, as it should be.}. The latter is also clear, 
since IDM relaxes to the standard dust at sufficiently late times. 
Hence, the non-adiabatic pressure is relevant only in the intermediate regime $t \simeq t_*$, 
when the effects due to the shift-symmetry breaking may become strong enough. 
They are encoded in the appearing Noether charge density in Eqs.~\eqref{epsnoether} and~\eqref{endens} and explicitly in terms 
involving the derivative of the function $\gamma$ in Eqs.~\eqref{pressure} and~\eqref{presspert}. 
These new terms source the non-adiabatic pressure. To summarize, an 
expected change induced in the curvature perturbation $\zeta_{IDM}$ is measured in terms of the 
quantities calculated at the times $t \simeq t_*$. Shortly, we will confirm this observation by making 
an exact calculation. We will also see that the variation of the curvature perturbation 
$\zeta_{IDM}$ is small, as it relies on the derivative of the potential $\Psi$.

To study the evolution of the curvature perturbation, Eq.~\eqref{evolution} is not very convenient. For this purpose, it is simpler to exploit Eq.~\eqref{nonconsnoether}. Integrating the latter 
and using Eq.~\eqref{back}, one obtains at the times $t>t_*$,
\begin{equation}
\label{c}
\delta n  =3 n  \left(\Psi (t_*)-\Psi \right) +2 n \cdot \frac{\dot{\Psi} (t_*)}{H(t_*)} \; .
\end{equation}
Using then Eqs.~\eqref{epsnoether},~\eqref{pressure},~\eqref{back},~\eqref{endens},~\eqref{00} and~\eqref{c}, we derive 
the expression for the curvature perturbation of IDM,
\begin{equation}
\label{synchr}
\zeta_{IDM} =\Psi +\frac{3n  (\Psi (t_*)-\Psi )+
6 n \cdot \frac{\dot{\Psi} (t_*)}{H(t_*)}+9 \gamma H \dot{\Psi}  }{3[n -3\gamma \dot{H}]} \; . 
\end{equation}
Hereafter, $\gamma$ denotes the value of the parameter at the end of the transition $\gamma_1 \rightarrow 
\gamma_2$, i.e., $\gamma \equiv \gamma_2$. Note that at the times $t<t_*$, when the Noether charge density equals to zero, 
Eq.~\eqref{synchr} reduces to the expression for the curvature perturbation of the 
total matter $\zeta_{tot}$, as it should be. Now, we are interested in the different regime, 
when IDM mimics the behaviour of the dust particles, i.e., strong inequalities $n \gg \gamma |\dot{H}| \sim \gamma H^2 \sim \gamma \rho_{tot}$ 
are obeyed. For the relevant values of the parameter $\gamma$ this still happens deeply in the RD stage. In that limit, we get for the curvature 
perturbation of IDM, 
\begin{equation}
\label{zetasynchr}
\zeta_{IDM}=\Psi (t_*) + 2\frac{\dot{\Psi} (t_*)}{H(t_*)} \; .
\end{equation}
The latter is constant as expected: the non-adiabatic pressure is negligible in the late-time regime. Now, let us show 
that the second term on the r.h.s. of Eq.~\eqref{zetasynchr} equals to zero, 
i.e., the derivative of the gravitational potential vanishes at $t=t_*$. This 
we will do in the reasonable approximation, when all the external matter is in the 
state of radiation\footnote{In particular, this statement is exact for the times $t_*$ corresponding 
to the temperatures $T \gtrsim 100$ GeV, when all the Standard Model 
degrees of freedom are relativistic.}. We will need the $ij$-component of Einstein's equations, which reads in the synchronous gauge,  
\begin{equation}
\label{ij}
\ddot{\Psi}+3H \dot{\Psi} =-\frac{1}{2} \left( \delta {\cal P}_{ext} +\delta {\cal P}_{IDM} \right) \; .
\end{equation}
For the relativistic external matter, one has $\delta {\cal P}_{ext}=\frac{1}{3} \delta \rho_{ext}$. 
Using this, we combine Eqs.~\eqref{00} and~\eqref{ij} to exclude the quantities describing the external matter, 
\begin{equation}
\nonumber 
\ddot{\Psi} +4H \dot{\Psi} =\frac{1}{6} \delta \rho_{IDM} -\frac{1}{2} \delta {\cal P}_{IDM} \; .
\end{equation}
We substitute expressions~\eqref{endens} and~\eqref{presspert} for the IDM energy density and pressure perturbations and rewrite the equation above as follows, 
\begin{equation}
\nonumber 
\frac{d}{dt}\left[\left(1-\frac{3\gamma}{2} \right)\dot{\Psi}  \right]+4H \left(1-\frac{3\gamma}{2} \right) \dot{\Psi} =\frac{\delta n}{6} \; .
\end{equation}
Integrating this out, we get 
\begin{equation}
\label{solu}
\left(1-\frac{3\gamma}{2} \right)\dot{\Psi} a^4=\frac{1}{6}\int^t_{t_i} a^4 (\tilde{t}) \delta n (\tilde{t}) d\tilde{t} +C \; . 
\end{equation}
Here $t_i <t$ and $C$ are the arbitrary constants. Recall now that the Noether charge density 
is zero at the times $t<t_*$. As we are interested in the behaviour of the potential $\Psi$ at 
$t=t_*$ and since the quantity 
$n$ is always finite, the first term 
on the r.h.s. of Eq.~\eqref{solu} vanishes. Hence, the solution for the derivative of the gravitational 
potential reduces to
\begin{equation}
\label{sol}
\dot{\Psi} (t_*)=\frac{C}{\left(1-\frac{3\gamma (t_*)}{2} \right)a^4 (t_*)} \; .
\end{equation}
For the arbitrary value of the constant $C$, this solution is discontinuous, as it follows from the 
behaviour of the parameter $\gamma$ at $t=t_*$. Note, however, that the r.h.s. of Eq.~\eqref{sol} 
translates into the decaying mode of the potential $\Psi$, which is commonly dropped in cosmology\footnote{Recall that the scale factor during the RD stage grows as $a \propto \sqrt{t}$. 
Hence, the decaying mode of the potential $\Psi$ drops as $\Psi_{dec} \propto \frac{C}{t}$.}. 
That is, we should set the constant $C$ to zero. In this situation, the derivative of the potential $\Psi$ vanishes, i.e., $\dot{\Psi}(t_*)=0$, and the expression for the 
IDM curvature perturbation~\eqref{zetasynchr} simplifies to $\zeta_{IDM}=\Psi (t_*)$.

Consequently, the potential $\Psi $ is a continuous constant function at the time $t=t_*$. From Eq.~\eqref{c}, we then conclude that $\delta n(t_*)=0$. The same is true for the IDM and external matter energy density perturbations, i.e., $\delta \rho_{IDM} (t_*)=0$ and $\delta \rho_{ext} (t_*)=0$. These follow from Eqs.~\eqref{endens} and~\eqref{00}, respectively. In the situation, when all the degrees of freedom 
are relativistic and there is no admixture of the isocurvature mode in the 
particle sector, one has an equality $\delta_{ext}=\delta_{ph}$, 
where $\delta_{ext} \equiv \frac{\delta \rho_{ext}}{\rho_{ext}}$ and $\delta_{ph} \equiv \frac{\delta \rho_{ph}}{\rho_{ph}}$; the subscript $''\mbox{ph}''$ stands for the photons. Hence, $\delta_{ph} (t_*)=0$. 
Recall now the expression for the curvature perturbation of the photons, 
\begin{equation}
\nonumber
\zeta_{ph} =\Psi +\frac{1}{4}\delta_{ph} \; .
\end{equation}
The latter stays constant behind the horizon, i.e., $\zeta_{ph} (t)=\zeta_{ph} (t_*)$. 
Thus, $\zeta_{ph}=\Psi (t_*)$. To summarize, there is 
no admixture of the IDM isocurvature mode, i.e.,
\begin{equation}
\label{final}
S_{IDM, ph}\equiv 3(\zeta_{IDM}-\zeta_{ph}) =0\; .
\end{equation}
Perturbations of IDM remain adiabatic after the shift-symmetry breaking, provided 
only that we can neglect the non-relativistic degrees of freedom during the transition $\gamma_1 \rightarrow \gamma_2$.

The result~\eqref{final} can be also understood from a slightly different prospective. As we noted 
earlier, the IDM curvature perturbation starts from exact adiabatic initial conditions~\eqref{adiabat} 
and may change only due to the non-zero adiabatic pressure~\eqref{nonadpr}. The latter can be relevant only 
at the times $t \simeq t_*$. Next, we observe that both the energy density perturbation (the numerator in Eq.~\eqref{synchr}) and 
the pressure perturbation~\eqref{presspert}, rely only on the variation of the potential $\Psi$. Accordingly to Eq.~\eqref{nonadpr}, so does the non-adiabatic pressure and, consequently, the 
isocurvature perturbation. Hence, the resultant perturbations must be adiabatic, once 
the potential $\Psi$ is constant at the relevant times.

Finally, let us comment on the generality of the results obtained. 
First, we notice that the assumption of the instantaneous transition 
$\gamma_1 \rightarrow \gamma_2$ is not a strong one at all. Our results hold for the different choices of the 
function $\gamma (\varphi)$ given that the 
gravitational potential $\Psi$ remains constant during the phase of the shift-symmetry breaking. 
That condition is satisfied with a high accuracy provided that the transition 
occurs well within the RD stage. 

Second, instead of the HD term as in Eq.~\eqref{action}, one could consider another one, 
\begin{equation} 
\label{hdother}
+\frac{\tilde{\gamma}(\varphi)}{2} \nabla_{\mu} \nabla_{\nu} \varphi \nabla^{\mu} \nabla^{\nu} \varphi \; .
\end{equation}
We provide the associated analysis in Appendix B. In particular, we show that all the 
three statements take place: i) in the shift-symmetric case, the equation of state of IDM is that of 
the total matter; 2) perturbations of IDM are exactly adiabatic in this situation; 
3) an approximate adiabaticity holds after the short phase of shift symmetry breaking. 

To conclude, initial conditions for IDM are the same as in the 
CDM case. Hence, CDM and IDM result with the same predictions 
regarding the cosmological observations. An important difference, 
however, may arise at the galaxy scales. 
This is due to the fact that IDM possesses non-zero sound speed. 
Furthermore, IDM and CDM exhibit an apparently different behaviour in the 
non-linear regime~\cite{Capela:2014xta}. We leave this and other interesting questions for the future. 

{\it {\bf Acknowledgments:}} We are indebted to Maxim Pshirkov and Sergey Sibiryakov for many useful discussions. 
We are particularly grateful to Alexander Vikman for the original idea and for the collaboration at the initial stages of the project.
This work is supported by the Belgian Science Policy IAP VII/37.


\section*{Appendix A}
In this Appendix, we switch to the Newton's gauge in order to cross-check the results obtained 
in the main body of the paper. In that case, the perturbation of the Noether charge density is given by 
\begin{equation}
\nonumber
\delta n=\delta \lambda+6\gamma \dot{H}\Phi +3\gamma H \dot{\Phi} -3\gamma \ddot{\Psi} \; ,
\end{equation}
where the shift-symmetry is assumed. 
The expression for the energy density perturbation of IDM reads in the Newton's gauge, 
\begin{equation}
\label{Newtonenergy} 
\delta \rho_{IDM}=\delta n -9\gamma H^2 \Phi +9\gamma 
H \dot{\Psi} \; .
\end{equation}
Recall that the Noether charge density equals to zero before the shift-symmetry 
breaking takes place. Integrating Eq.~\eqref{nonconsnoether} with the initial 
condition $n=0$ set at the times $t<t_*$, one obtains at the times $t>t_*$,
\begin{equation}
\label{Newtonshift}
\delta n=-2 n \cdot \Phi (t_*)+
3 n  \cdot (\Psi 
(t_*)-\Psi) +\frac{n \delta \varphi (t_*)}{H(t_*)} \cdot {\cal P}_{tot} (t_*) +2 n \cdot 
\frac{\dot{\Psi} (t_*)}{H(t_*)} \; ,
\end{equation}
where $n\cdot a^3=\mbox{const}$ is defined by Eq.~\eqref{back}. Substituting Eq.~\eqref{Newtonshift} into Eq.~\eqref{Newtonenergy}, and then the latter into~\eqref{zeta}, 
we obtain for the IDM curvature perturbation, 
\begin{equation}
\label{Newtoncurv}
\zeta_{IDM} =\Psi + \frac{1}{3[n-3\gamma \dot{H}]} \times \Bigl[ 3n (\Psi (t_*) -\Psi )+n \cdot \delta_{tot} (t_*)
+\frac{3}{2} \gamma \cdot \delta \rho_{tot}+\frac{n \cdot \delta \varphi (t_*)}{H(t_*)} \cdot {\cal P}_{tot} (t_*)\Bigr] \; ,
\end{equation}
where $\delta_{tot} \equiv \delta \rho_{tot}/\rho_{tot}$. Here we made use of the $00$th component of Einstein's equations, which reads in the Newton's gauge (in the super-horizon regime), 
\begin{equation}
\nonumber 
3H \dot{\Psi} -3H^2 \Phi=\frac{1}{2} \delta \rho_{tot} \; .
\end{equation}
The curvature perturbation~\eqref{Newtoncurv} corresponds to exactly adiabatic initial conditions for IDM before the Noether charge is produced, as expected. 
Well after the shift-symmetry gets broken, i.e., 
in the regime $n \gg \gamma H^2$, one has
\begin{equation}
\nonumber 
\zeta_{IDM}=\Psi (t_*) +\frac{1}{3} \delta_{tot} (t_*) +\frac{1}{3} \frac{\delta \varphi (t_*)}{H(t_*)} \cdot {\cal P}_{tot} (t_*) \; .
\end{equation}
This expression is equal to the curvature perturbation~\eqref{zetasynchr} calculated in the 
synchronous gauge, as it should be. In particular, neglecting the super-horizon variation of the potential $\Psi$ at the time $t_*$ 
and using $\delta \varphi \approx \Phi t$ and $H \approx \frac{1}{2t}$, we get
\begin{equation}
\nonumber
\zeta_{IDM} \approx \Psi (t_*) +\frac{\delta \rho_{tot} (t_*)}{4 \rho_{tot} (t_*)} \approx \zeta_{ph} \; .
\end{equation}
This corresponds to the adiabatic initial conditions for IDM.

\section*{Appendix B}

In this Appendix we consider another 
possible HD term given by Eq.~\eqref{hdother}. The associated energy momentum tensor is given by~\cite{Haghani:2014ita}  
\begin{equation}
\nonumber 
T_{\mu \nu} =\lambda \partial_{\mu} \varphi \partial_{\nu} \varphi +\tilde{\gamma} \nabla^{\lambda} ( \nabla_{\lambda} \varphi \nabla_{\mu} \nabla_{\nu} \varphi ) 
-\tilde{\gamma} \square \nabla_{\mu} \varphi \nabla_{\nu} \varphi-\tilde{\gamma} \square \nabla_{\nu} \varphi \nabla_{\mu} \varphi-\frac{1}{2} \tilde{\gamma} g_{\mu \nu} \nabla^{\alpha} \nabla^{\beta} \varphi 
\nabla_{\alpha} \nabla_{\beta} \varphi \; .
\end{equation} 
At this level, we assumed the shift-symmetry, as it is going to be enough for our purposes. 

First, it is straightforward to show that in an exactly shift-symmetric case, 
IDM still tracks the dominant matter of the Universe. Indeed, from the energy-momentum tensor one deduces for the 
energy density and the pressure, 
\begin{equation}
\label{rhoother}
\rho_{IDM}=n +\frac{\tilde{\gamma}}{2} \rho_{tot}
\end{equation}
and 
\begin{equation}
\label{pressother}
{\cal P}_{IDM}=\frac{\tilde{\gamma}}{2} {\cal P}_{tot} \; .
\end{equation}
Here $n$ is the Noether charge density now given by the expression 
\begin{equation}
\nonumber 
n \equiv J^{0}= \lambda \partial^0 \varphi - \nabla^{\nu} \left(\tilde{\gamma} \nabla_{\nu} \nabla^{0} \varphi \right)
\end{equation}
As it follows, in the situation, when the Noether charge is zero, the 
equation of state of IDM is that of the total matter in the Universe. 
Not surprisingly thus, super-horizon modes of IDM are exactly of the 
adiabatic type.

To show this explicitly, we set to zero the Noether charge density $n$ as well as its super-horizon perturbation, 
\begin{equation}
\nonumber
\delta n =\delta \lambda +6\tilde{\gamma} H\dot{\Psi} \; .
\end{equation}
The relation between the IDM energy density perturbation and the quantity $\delta n$ is given by 
\begin{equation}
\label{rhoappb}
\delta \rho_{IDM} =\delta n +3\tilde{\gamma} H \dot{\Psi} \; .
\end{equation}
Hence, 
\begin{equation}
\nonumber 
\delta \rho_{IDM} =3\tilde{\gamma} H \dot{\Psi} \; .
\end{equation}
Substituting this into the definition~\eqref{zeta}, using Eqs.~\eqref{rhoother} and~\eqref{pressother}, one obtains $\zeta_{IDM}=\zeta_{tot}$. The latter implies an exact adiabaticity of IDM 
perturbations in the situation, when the Noether charge density equals to zero.

To obtain the expression for the perturbation $\zeta_{IDM}$ in the generic case, one exploits an equation, 
\begin{equation}
\label{noetherother}
\frac{d}{dt} \left(\sqrt{-g} \cdot n   \right) =\frac{\tilde{\gamma}_{\varphi}}{2}  \sqrt{-g} \nabla_{\mu} \nabla_{\nu} \varphi \nabla^{\mu} \nabla^{\nu} \varphi \; .
\end{equation}
Namely, we promote the constant $\tilde{\gamma}$ to the function of the field $\varphi$. 
We again assume with no loss of generality, the instantaneous transition of the parameter 
$\tilde{\gamma}$ as in Eq.~\eqref{inst}. Integrating Eq.~\eqref{noetherother} over the time, we get
\begin{equation}
\nonumber 
n\cdot a^3 =\frac{3}{2}\Delta \tilde{\gamma} H^2 a^3 \left.\right|_{t=t_*} \qquad t>t_*\; .
\end{equation}
The analogous equation at the level of super-horizon fluctuations coincides with Eq.~\eqref{c}.
Substituting Eq.~\eqref{c} into Eq.~\eqref{rhoappb}, we obtain for the IDM energy density perturbation
\begin{equation}
\label{resultother}
\delta \rho_{IDM} =3n \cdot (\Psi (t_*)-\Psi )+n \cdot \delta_{tot}(t_*)+
\frac{1}{2}\tilde{\gamma} \cdot \delta \rho_{tot}  \; .
\end{equation} 
We substitute this into Eq.~\eqref{zeta}. At sufficiently late times, i.e., when the 
strong inequality $n \gg \tilde{\gamma} \rho_{tot}$ is obeyed, we arrive at the expression~\eqref{zetasynchr}. Again 
neglecting the super-horizon variation of the potential $\Psi$ and then following 
the same steps as in the bulk of the paper, one concludes with adibaticity of 
IDM perturbations.

\end{document}